\documentclass[preprint2]{aastex631}

\shorttitle{The Radius Cliff}
\shortauthors{Dattilo \& Batalha}
\graphicspath{{./}{figures/}}

\usepackage{xspace}

\newcommand{\Kepler}{\textit{Kepler}\xspace}
\renewcommand{\Re}{R$_\oplus$\xspace}
\newcommand{\Si}{S$_\oplus$\xspace}

\begin{document}

\title{A Unified Treatment of Kepler Occurrence to Trace Planet Evolution II: The Radius Cliff Formed by Atmospheric Escape}

\author[0000-0002-1092-2995]{Anne Dattilo}
\affiliation{Department of Astronomy and Astrophysics, University of California Santa Cruz, Santa Cruz, CA, 95064, USA}

\author[0000-0002-7030-9519]{Natalie M. Batalha}
\affiliation{Department of Astronomy and Astrophysics, University of California Santa Cruz, Santa Cruz, CA, 95064, USA}

\correspondingauthor{Anne Dattilo}
\email{adattilo@ucsc.edu}

\begin{abstract}

The Kepler mission enabled us to look at the intrinsic population of exoplanets within our galaxy. In period-radius space, the distribution of the intrinsic population of planets contains structure that can trace planet formation and evolution history. The most distinctive feature in period-radius space is the radius cliff, a steep drop-off in occurrence between $2.5-4$\Re across all period ranges, separating the sub-Neptune population from the rarer Neptunes orbiting within 1~au. Following our earlier work to measure the occurrence rate of the Kepler population, we characterize the shape of the radius cliff as a function of orbital period ($10-300$ days) as well as insolation flux (9500\Si -- 10\Si). 
The shape of the cliff flattens at longer orbital periods, tracking the rising population of Neptune-sized planets. In insolation, however, the radius cliff is both less dramatic and the slope is more uniform. The difference in this feature between period- and insolation-space can be linked to the effect of EUV/X-ray versus bolometric flux in the planet's evolution.
Models of atmospheric mass loss processes that predict the location and shape of the radius valley also predict the radius cliff. We compare our measured occurrence rate distribution to population synthesis models of photoevaporation and core-powered mass-loss in order to constrain formation and evolution pathways. We find that the models do not statistically agree with our occurrence distributions of the radius cliff in period- or insolation-space. Atmospheric mass loss that shapes the radius valley cannot fully explain the shape of the radius cliff.

\end{abstract}

\section{Introduction} \label{sec:intro}
\Kepler's foremost legacy has been enabling detailed exoplanet demographic work. With a catalog of over 5,000 planet candidates, we can begin to understand the intrinsic population of planets and the mechanisms that form them. One of the first significant results in exoplanet demographics was the observation of the radius valley. Several theories had previously predicted this bimodal feature as a byproduct of photoevaporation \citep{lopez2013, owen2017, lopez2018}. Only after carefully measuring stellar and planet properties could it be seen in the observed planet occurrence \citep{fulton2017}. 
Other examples of theoretical predictions that have been verified by observational measurements include the giant planet turnover past the ice line \citep{ida2004, fernandes2019}.
We can constrain planet formation and evolution mechanisms by studying the intrinsic population of planets in the galaxy.

\begin{figure}
    \centering
    \includegraphics[width=\columnwidth]{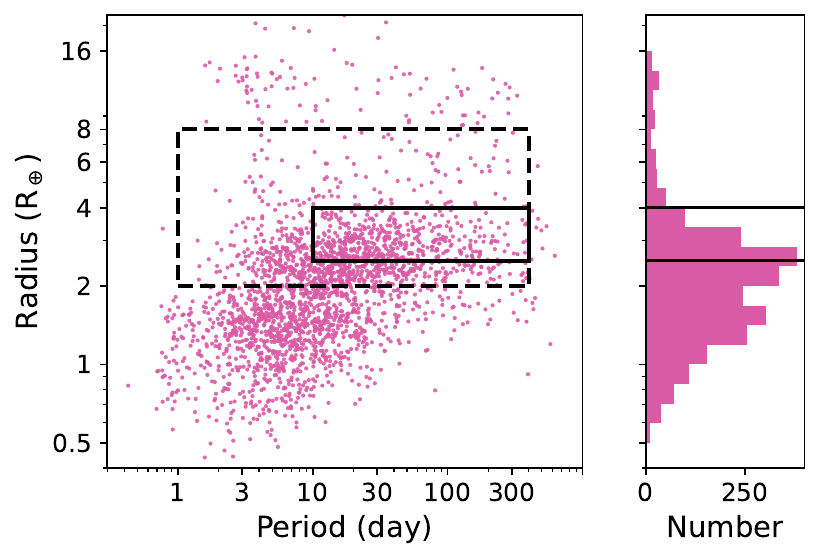}
    \caption{Period-radius diagram of the DR25 \Kepler planet candidate sample on the left and a histogram of the observed planet radii on the right. The larger dashed box denotes the region we are using for analysis, while the smaller solid box denotes our definition of the radius cliff.}
    \label{fig:prdiagram}
\end{figure}

However, few works in the literature have focused on the most notable feature in the period-radius diagram: ``radius cliff,'' a steep drop in occurrence that separates the ubiquitous sub-Neptune-sized planets from the rarer Neptune-sized ones. Notably, this feature is seen in both the observed planet distribution and the survey-completeness corrected occurrence distribution. We define the region of the radius cliff to be 2.5-4.0 \Re and from 10-300 days, marked in the inner box in Figure~\ref{fig:prdiagram}, as this is past the dearth of the Neptune desert and extends in radius from the peak of the small planet distribution.

\Kepler{} RV and TTV follow-up has shown that sub-Neptunes are low-density and cannot be entirely rocky but require gaseous envelopes \citep{weiss2014, hadden2014}. 
We use the term ``sub-Neptune'' throughout this work to mean planets with an Earth-like core and a large H/He atmosphere.
These planets are mainly core by mass, while their radius is dominated by their atmosphere. To explain the over-abundance of sub-Neptunes and lack of Neptunes-- i.e., to explain the existence of the radius cliff-- the planets' atmospheric growth must be inhibited by some physical or formation process. While sub-Neptunes are the most common type of planet in our galaxy, we have no Solar System equivalent, making it difficult to understand their formation and evolution processes.

Several planet evolution theories create the radius cliff as a secondary outcome to other period-radius diagram features. Mass loss via photoevaporation of the planet's atmosphere form the radius valley and explain an abundance of sub-Neptunes at short orbital period \citep{owen2017}. Core-powered mass-loss, used as a primary explanation for the radius valley, also creates an abundance of sub-Neptunes at short orbital periods and a radius cliff \citep{ginzburg2018}. 

While a radius cliff-like feature is produced in each of these models, they are developed to match the slope of the radius valley, not the observed location and occurrence of the cliff. If the models well-reproduce the radius cliff of the \Kepler sample, that model alone could explain the evolutionary history of the sub-Neptune population. 


In this paper, we use the methods of \citet{dattilo2023} (described in Section~\ref{sec:methods}) to characterize the shape of the occurrence cliff as a function of orbital period and insolation flux (Section~\ref{sec:occ}). We then compare the shape of the cliff with two atmospheric mass loss models to see if they can accurately explain the cliff shape (Section~\ref{sec:formation}). Finally, we offer alternative mechanisms in Section~\ref{sec:discussion} and conclude in Section~\ref{sec:conclusions}.

\section{Stellar \& Planet Sample} \label{sec:stars}
We use the same stellar sample of FGK stars as \citet{dattilo2023} and briefly describe the sample here.
We base our stellar sample on the \Kepler DR25 catalog \citep{mathur2017} and use the Gaia-derived parameters from \cite{berger2020a}. This catalog has improved stellar radii, which correspondingly improves the accuracy of our planet radii. We make various quality cuts, described in detail in \citet{dattilo2023}, that result in a stellar sample of 82,912 FGK stars. Using the temperature boundaries from \citet{pecaut2013}, we split the sample into 21,999 F stars $(6000$ K $\leq T_{eff} < 7300$ K); 41,501 G stars $(5300$ K $\leq T_{eff} < 6000$K); 19,412 K stars $(3900$ K $\leq T_{eff} < 5300$ K).

We draw our planet sample from the Q1-17 DR25 KOI table, restricted to KOIs with \texttt{koi\_disposition} = CANDIDATE on the stars in our parent stellar sample. We recompute the planet radius, $R_P$, based on the ratio of planet radius to stellar radius, $R_P / R_*$ (\texttt{koi\_ror}), using the updated stellar radii from \citet{berger2020a}. We further restrict this sample to the period and radius range $1-400$ days and $1-10$\Re to calculate occurrence (dashed box in Figure~\ref{fig:prdiagram}). There are 479 planets around F stars, 650 planets around G stars, and 695 planets around K stars, for a total of 2124 planets in our FGK sample.

\section{Methodology}\label{sec:methods}
To calculate occurrence, we use the methodology from \citet{dattilo2023} and briefly recap it here:

We use a kernel density estimator (KDE) methodology to measure the distribution of observed exoplanets in the \Kepler sample. We include completeness and reliability measurements to account for survey bias and false positives. Details of how reliability and completeness are computed are detailed in \citet{bryson2020a}. The KDE is computed by placing a 2D Gaussian kernel at each planet's respective period and radius with size $(\sigma_p, \sigma_r) = (0.1, 0.1)$. Planet reliability is included in this measurement as the amplitude of the respective planet kernel in period-radius space (equivalent to its Gaussian volume). We compute a per-star completeness contour, including detection efficiency, geometric transit probability, and vetting efficiency,  via a modified version of KeplerPORTS \citep{burke2017, bryson2020a}. Total completeness is computed as the sum of the per-star completeness contour for all stars in the sample. We measure occurrence by dividing the KDE map by the completeness contour. Our uncertainties are computed by bootstrapping the sample 5000 times within their radius errors.

\subsection{Converting to Insolation-Radius Space}\label{sec:insol_method}
For this work, we also compute occurrence in insolation-radius space. We calculate the insolation of each planet based on their stellar properties from the \citet{berger2020a} Gaia-\Kepler catalog. The insolation of each planet, $S$, is defined by:

\begin{equation}\label{eq:insol}
    \frac{S}{S_\oplus} = \left(\frac{R_*}{R_\odot}\right)^2 \left(\frac{T_{eff}}{T_\odot}\right)^4 \left(\frac{M_*}{M_\odot}\right)^\frac{2}{3} \left(\frac{P}{\textrm{yrs}}\right)^{-\frac{4}{3}}
\end{equation}

where $R_*$, $T_{eff}$ and $M_*$, are the stellar radius, effective temperature, and mass pulled from \citet{berger2020a} and planet orbital period, $P$, is pulled from the DR25 catalog. This gives each planet a period, radius, insolation, and respective uncertainties. 

Using the same method as for period-radius space, we compute insolation-radius occurrence by creating a KDE of the observed population of exoplanets and dividing it by the computed insolation completeness contour.

\subsubsection{Insolation Completeness}
To compute the completeness for insolation-radius space, we take the per-star completeness contour in period-radius space and transform period to insolation via Equation~\ref{eq:insol}, following \citet{bryson2021}.

Because each stellar type bin includes a range of effective temperatures, a given period leads to disparate insolation fluxes, even for stars of the same stellar type. The \Kepler completeness function is only valid to 500 days, and this is equivalent to a range of insolation fluxes for different stars. To measure occurrence, we need to define a range in insolation that does not exceed a 500 day orbital period for all stars in their respective stellar type bin.



The maximum insolation value for each stellar type bin is computed as the minimum of the set of fluxes at the surface of the star. The minimum insolation value is computed as the maximum insolation at 500 days. By setting these boundaries, we remove a large area of parameter space but ensure we make the same assumptions about all stars. For FGK stars, this range is 9500 -- 12 \Si; for F stars, this range is 50,000 -- 12 \Si; for G stars this is 30,000 -- 5\Si; and for K stars, this is 9500 -- 2\Si.

\section{Occurrence Measurements}\label{sec:occ}
\subsection{Period-Radius Occurrence}\label{sec:period}
\begin{figure*}[ht!]
    \centering
    \includegraphics[width=\textwidth]{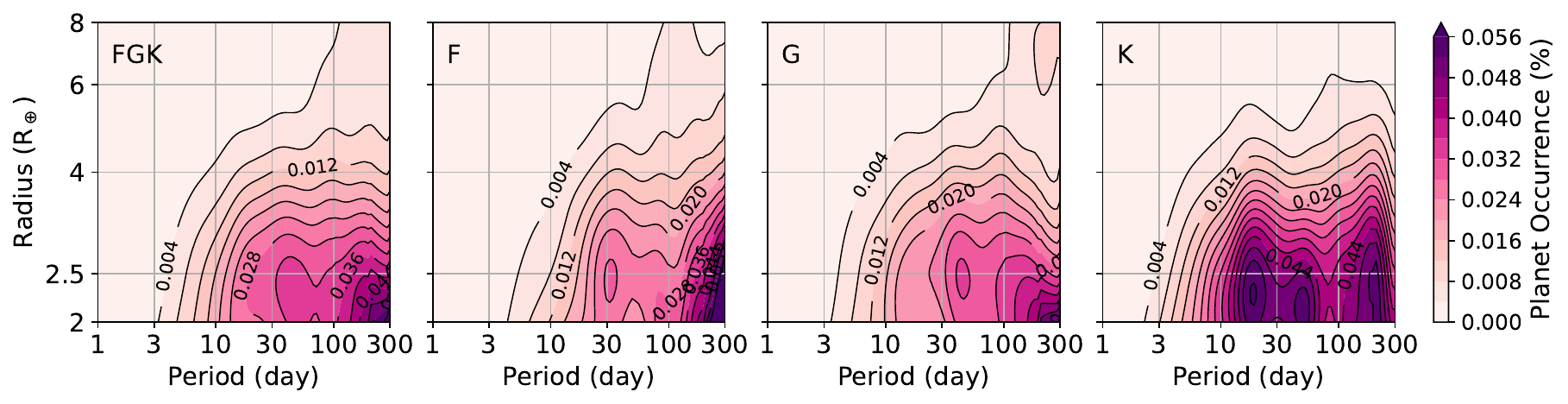}
    \caption{Occurrence contours in period-radius space. The radius range only shows the sub-Neptune and sub-Jupiter range of planets, from $2-8$\Re. Each plot shows occurrence as a percentage for different stellar types: FGK, F, G, K, respectively.}
    \label{fig:2D_period}
\end{figure*}

We measure the occurrence distribution from the sub-Neptune population to below the hot Jupiter population, $2-8$\Re. In the leftmost panel of Figure~\ref{fig:2D_period}, the peak of the FGK sub-Neptune population lies between $2-2.5$\Re. The radius valley lies below the radius range of the plots and is not shown. Above 2.5\Re, occurrence quickly drops off in what we call the ``radius cliff''. 

Occurrence is not uniform across all orbital periods. In the left side of the period-radius plane, planet occurrence is very low; this is the Neptune desert. The edge of the desert wraps around the sub-Neptune population and can be seen via the lowest occurrence contours. Sub-Neptunes become increasingly common at longer orbital periods, shown by the rising occurrence past 100 days for 2--2.5\Re.

The occurrence distribution also varies as a function of stellar type, as seen in the three right-hand panels in Figure~\ref{fig:2D_period}. Qualitatively, the edge of the Neptune desert lies at shorter orbital periods with later-type stars.
The shape of the edge of the Neptune desert also changes across stellar types, with G- and K-type stars having more sub-Neptunes between 10--30 days, giving the desert a ``rounder'' edge. Also, the radius cliff extends to larger radii for F- and G-type occurrence than it does for K-type occurrence.

The Neptune desert and the radius cliff are a single continuous feature that delineates the largest sub-Neptunes from the Neptunes. The cliff edge and radius cliff, within a few hundred day orbital period, are likely shaped by the same physical process (or processes). The contours in Figure~\ref{fig:2D_period} show that the shape of the cliff is not uniform in radius or period. How the shape of the cliff varies will tell us what physics dominates those planets.


\subsubsection{Radius Cliff Slope} \label{sec:period-cliff}

In order to characterize the shape of the radius cliff, we follow the methodology of \citet{dattilo2023} and measure the slope of the cliff for different period ranges. By marginalizing the cliff into a radius distribution, we can quantify the change in steepness of the cliff as a function of orbital period.

\begin{figure}
    \centering
    \includegraphics[width=\columnwidth]{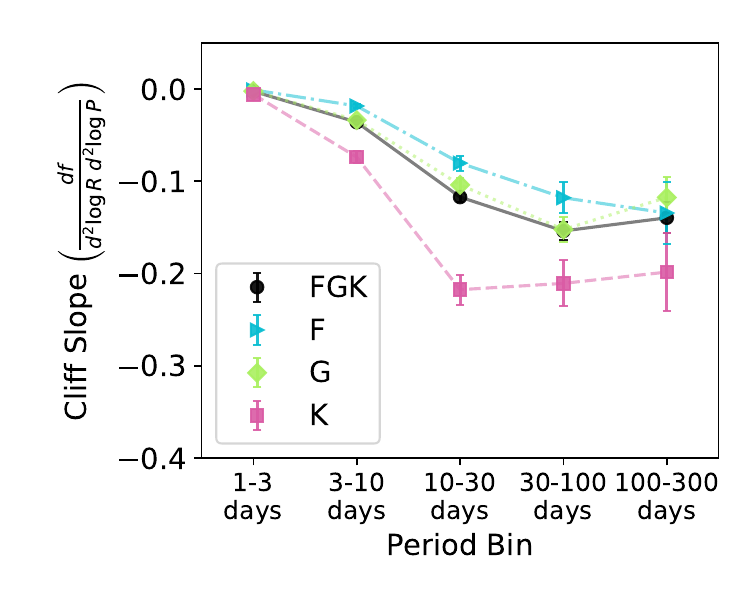}
    \caption{The slope of the radius cliff for each stellar type bin. The slope of the cliff is measured as in Equation~\ref{eq:cliff}.}
    \label{fig:per_slopes_dType}
\end{figure}

To measure the slope, we take a marginalized radius distribution for log-uniform period bins, with edges: $P = [1, 3, 10, 30, 100, 300]$ days. We then measure the slope of the cliff, $m_C$, by fitting the function:
\begin{equation}\label{eq:cliff}
    \textrm{NPPS} = F = m_C \log{R_{P}} + b
\end{equation}
where $F$ is occurrence in Number of Planets Per Star (NPPS) and $b$ is the intercept of the fit line which acts as a nuisance parameter. We fit this equaation for the radius range $2.5-6$\Re. This is equivalent to $\frac{df}{d^2 \log R\; d^2 \log P}$. We plot the fit slopes in Figure~\ref{fig:per_slopes_dType} for each stellar type bin.


At short orbital periods, the radius cliff becomes steeper as orbital period increases. 
For the FGK sample, the cliff is steepest between $30-100$ days, with the next period bin, $100-300$ days, having a similar slope. Looking at Figure~\ref{fig:2D_period}, the similar cliff slope measurements between 100--300 days may be surprising because the sub-Neptune occurrence rises very quickly. However, the slope is measuring the \textit{relative} amount of sub-Neptunes to Neptune- and Saturn-sized planets. The occurrence of the larger planets proportionally increases in the 100-300 day bin, keeping the cliff slope near to the 30-100 day slope.
The contours of Figure~\ref{fig:2D_period} show this rise in larger planet occurrence, and therefore the changing slope of the cliff, via the lowest occurrence contours. Instead of staying at a constant radius, the lowest contour extends to larger radii at longer orbital periods.

The trend is not the same for separate F-, G-, and K-type occurrence. For K-type stars, the cliff becomes quite steep as early as $10-30$ days, where sub-Neptune occurrence stays relatively constant to 300 days. G-type occurrence is very similar to the FGK occurrence, and the steepest cliff slopes is in the 30--100 day bin. For F-type occurrence, it is unclear if we measure the steepest cliff slope within this period range.

The change in slope of the radius cliff, from zero at the smallest period bin to the steepest slope, traces out the edge of the Neptune desert.
The location of the steepest slope would mark the end of the Neptune desert as that is where Neptune and Saturn-sized planets become relatively more common.
This means that the edge of the Neptune desert moves inwards for later-type stars.






\subsection{Insolation-Radius Occurrence}\label{sec:insol}
Using the methods of \citet{bryson2020b} to transform \Kepler{} completeness into insolation-radius space (described in Section~\ref{sec:methods}), we can measure occurrence in insolation-radius space. 

\begin{figure*}
    \centering
    \includegraphics[width=\textwidth]{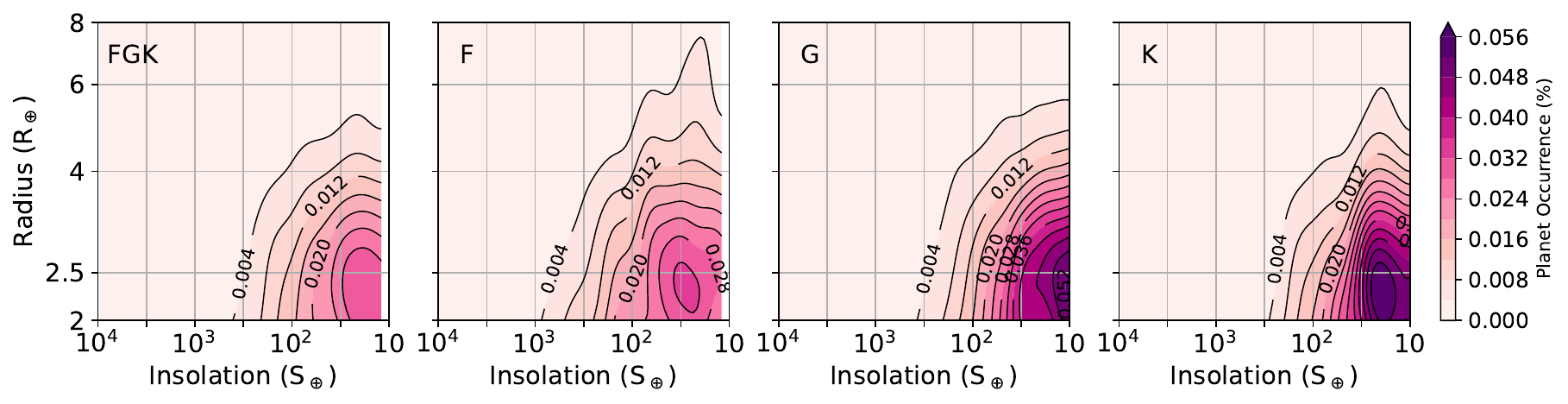}
    \caption{Occurrence contours in insolation-radius space. Note that the FGK and F occurrence plots have a lower insolation limit at 12\Si where the G and K occurrence plots have a lower insolation limit at 10\Si.}
    \label{fig:2D_insol}
\end{figure*}

We show occurrence contours for insolation-period space in Figure~\ref{fig:2D_insol}. The radius range is the same as in Figure~\ref{fig:2D_period}. We extend the insolation range out to $10^4$\Si because there are planets at those insolations, even though the occurrence is very close to zero. The different stellar type occurrences look comparatively similar to each other relative to their period-radius occurrence contours. This is expected because our stellar type bins are based on effective temperature, and have more relation to each planet's insolation flux than their orbital period. 

The Neptune desert is still apparent in these occurrence plots, with the radius valley not shown (as it's below 2\Re). The inner edge of the Neptune desert moves to lower insolation values for later type stars, as it does in period-space. In insolation, however, the radius cliff is not as apparent as it is in period-space. For period-radius occurrence, the period distribution obeys a broken power law with a steep rise in occurrence to 11.12 days, and then a flattening of the slope at long orbital periods \citep{dattilo2023}. The radius cliff occurs past the break in the broken power law. We do not measure the flat occurrence relation at low insolations. It is possible that we are only probing the edge of the desert and that the cliff-equivalent structure occurs at lower insolations. We still measure the cliff slope orthogonal to the insolation axis for Neptune desert comparisons.

There appears to be some period-dependent structure at $\sim$70 days in the occurrence plots of Figure~\ref{fig:2D_period} that are not apparent in the insolation occurrence plots in Figure~\ref{fig:2D_insol}. The dip in occurrence does not have high statistical significance. Insolation is the more physically-motivated property of planet formation, and we do not measure the same structure in insolation. The period-dependent peaks and valleys are likely due to the range of period values a single insolation flux value equates to for our range of stellar temperatures.


\subsubsection{Radius Cliff Slope}\label{sec:insol-cliff}
\begin{figure}
    \centering
    \includegraphics[width=\columnwidth]{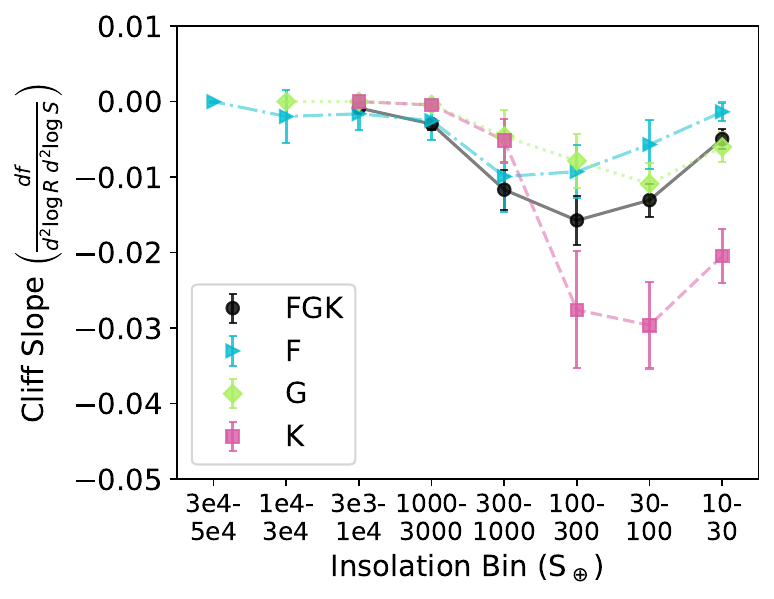}
    \caption{Slope of the radius cliff as a function of insolation flux. The slope of the cliff is calculated as Equation~\ref{eq:cliff}, where $d^2 \log P$ is replaced by $d^2 \log S$.}
    \label{fig:insol_slopes_dType}
\end{figure}

We also measure cliff slope as a function of insolation flux.
We use log-uniform insolation bins to measure the slope of the cliff with the same process as described in Section~\ref{sec:period}, with bin edges: $S = [3\times 10^4, 1\times 10^4, 3000, 1000, 300, 100, 30, 10] S_\oplus$. Figure~\ref{fig:insol_slopes_dType} compares the slopes for each stellar type within each insolation bin.

Unlike the orbital period dependence of the cliff slopes, there is not a clear dependence on stellar type when analyzed as a function of insolation flux, seen in Figure~\ref{fig:insol_slopes_dType}. The steepest slope is at the highest insolations for F-type stars, a reversal of the period-dependent slopes. K-type occurrence reaches the steepest slope next, then G-type occurrence. 
The stellar type dependence of the shape of the cliff is clear in period-- the cliff reaches its steepest slope and then flattens out at shorter periods for cooler stars. There is not a clear story in insolation. Because orbital period can be used as a proxy for X-ray flux received from the host star, and likewise insolation for bolometric flux, we can hypothesize that the cliff is primarily formed by X-ray photoevaporation.

\section{Formation of the Cliff}\label{sec:formation}

Theoretical models of the radius valley produce an upper radius limit to the sub-Neptune population that mimics the observed population. Here, we quantify how well these models replicate the intrinsic radius cliff distribution that we measure in occurrence.
We take two recent theories of atmospheric mass loss and generate synthetic populations of planets that have evolved through billions of years of their respective physics. By comparing the synthetic population to our occurrence distribution, we are comparing two ``intrinsic'' populations.


We compare the shape of the cliff in two ways: comparing the KDE contours to 1) a 1-dimensional Kolmogorov-Smirnoff (KS) test between radius distributions and 2) a 2-dimensional KS test in period-radius space. A KS test is a non-parametric test that two samples came from the same probability distribution \citep{fasano1987}. We use a 2D KS rather than a $\chi^2$ goodness-of-fit test as it is more powerful in retaining the two dimensional shape information from the samples. We also computed Anderson-Darling statistics, a similar CDF-comparison test, but the \texttt{scipy} implementation puts a lower limit on p-values at 0.001, so we do not include these values in our tables. For the KDE comparisons, we normalize the KDEs to have the same total integrated occurrence. The KS test compares populations, not distributions, so we generate a population of planets from our occurrence distribution by drawing planet periods and radii from a Poisson function with rate equal to the occurrence rate of its respective period-radius cell. This produces greater than 10x the number of simulated planets for our defined radius cliff parameter space. To minimize any errors due to the difference in sample sizes, we draw an equal number of occurrence-intrinsic planets to model planets and bootstrap that sample 1000 times to calculate the relevant statistic.

The two atmospheric mass loss models we compare occurrence to are a photoevaporation model from \citet{rogers2021} and a core-powered mass-loss model from \citet{gupta2020}. We independently replicated the models of both of theories and validated our code with public benchmarks in their papers.



\begin{figure*}
    \centering
    \includegraphics[width=\textwidth]{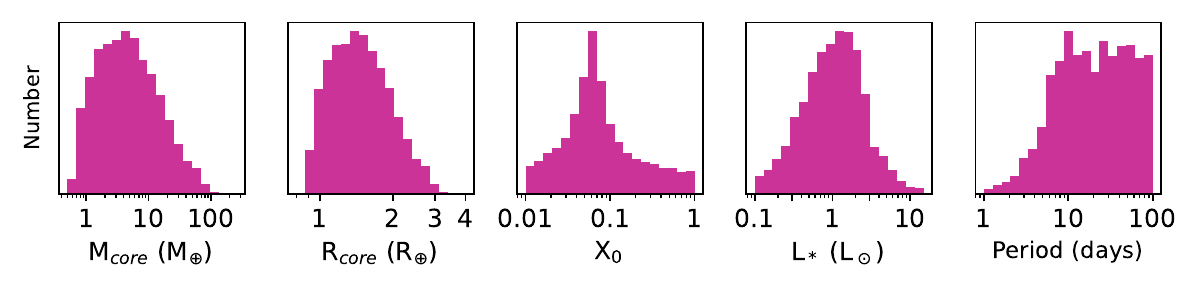}
    \includegraphics[width=\textwidth]{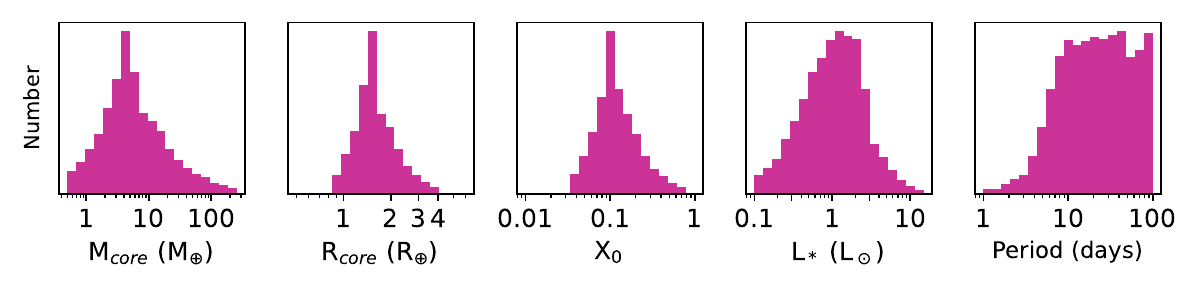}
    \caption{Input distributions for each population synthesis model. 
    Top: Model I, photoevaporation model from \citet{rogers2021}. 
    Bottom: Model II, core-powered mass-loss model from \citet{gupta2020}. 
    In each model, the input core mass, initial atmospheric mass fraction, and period distributions are taken for the best-fit parameters of their paper. The stellar distributions are taken from our \Kepler sample of stars with \textit{Gaia}-derived properties.}
    \label{fig:input-dists}
\end{figure*}

\begin{figure*}
    \centering
    \includegraphics[width=\textwidth]{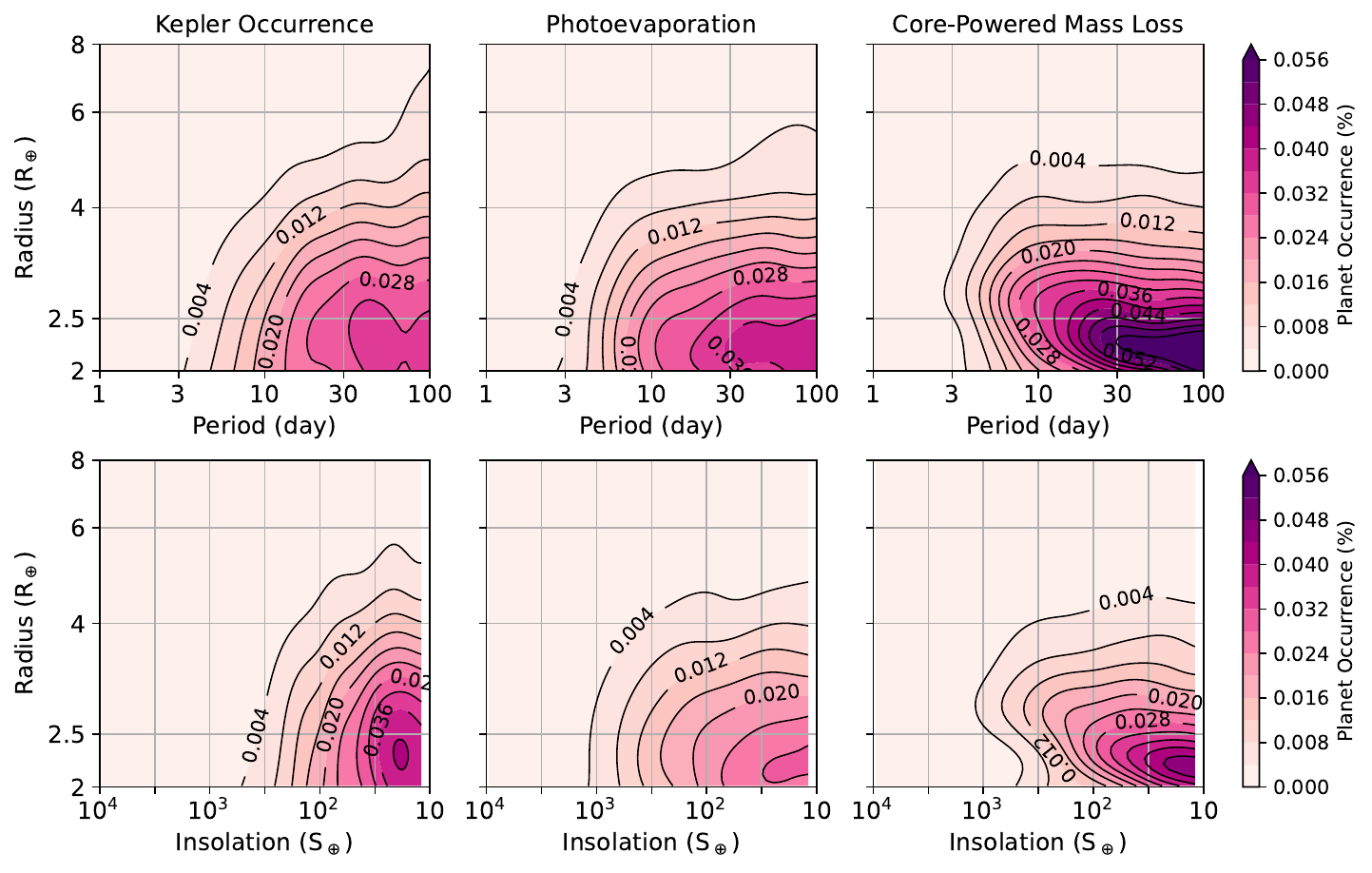}
    \caption{FGK occurrence comparisons with synthetic model populations.
    First row is occurrence in the period-radius plane, from 1-100 days and 2-8\Re.
    Second row is occurrence in the insolation-radius plane, from $10^4$S$_\oplus$-10 S$_\oplus$.}
    \label{fig:FGK-comp}
\end{figure*}

\begin{figure*}
    \centering
    \includegraphics[width=\textwidth]{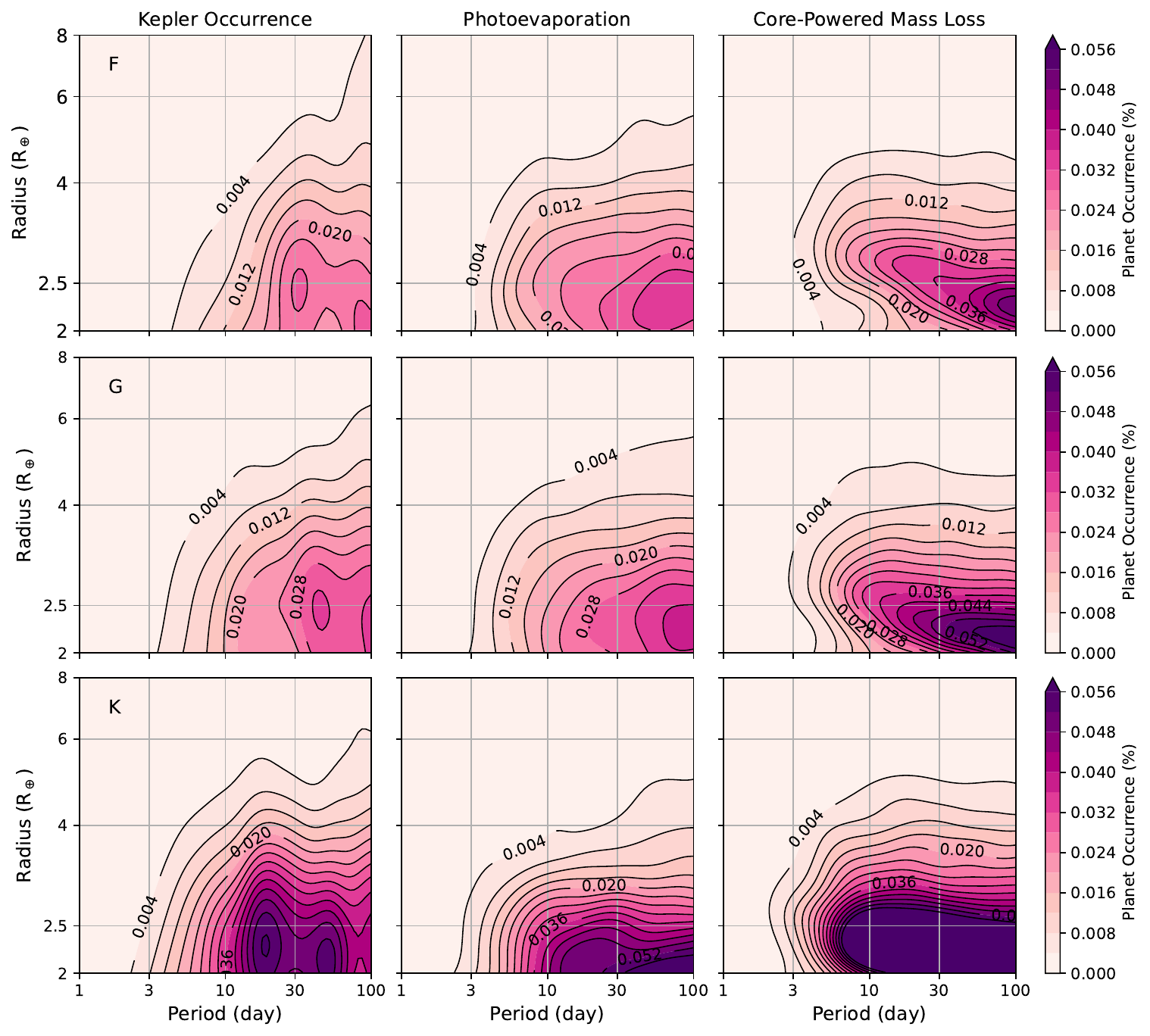}
    \caption{Occurrence-model comparisons for F-, G-, and K-type stellar type bins in period-radius space.}
    \label{fig:dTypes-comp}
\end{figure*}

\begin{figure*}
    \centering
    \includegraphics[width=\textwidth]{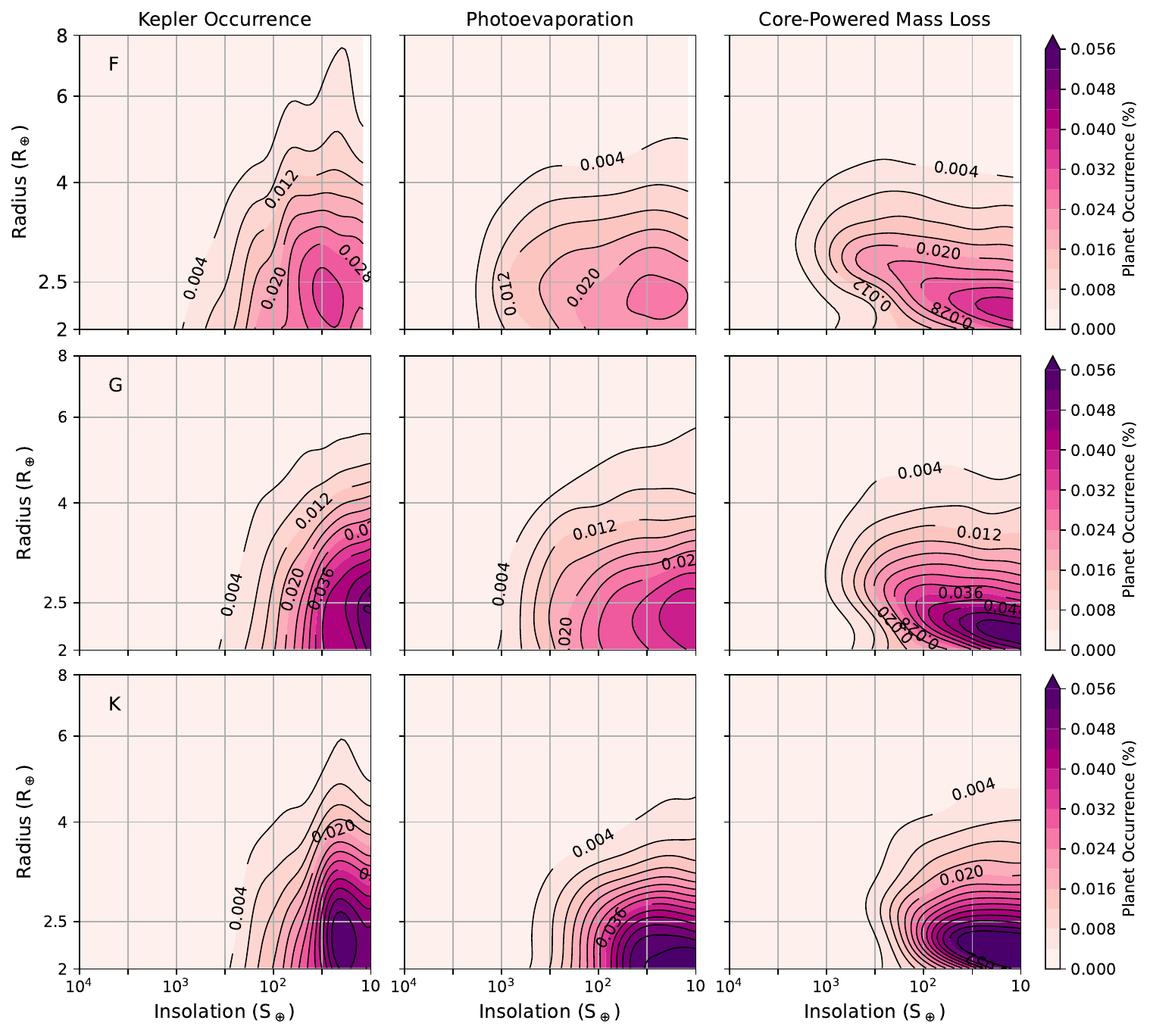}
    \caption{Occurrence-model comparisons for F-, G-, and K-type stellar type bins in insolation-radius space. The three models shown are \citet{rogers2021} and \citet{gupta2020}, respectively. Note that the contours for the K-type occurrence for core-powered mass-loss are truncated at 0.056\% for comparison with all other plots.}
    \label{fig:dTypes-comp-insol}
\end{figure*}

\subsection{Model I: Photoevaporation}\label{sec:photoevap}
Our photoevaporation model is based on the theoretical model presented in \citet{rogers2021}. We build our code based on the code of \citet{owen2020}.
For the input distributions, we draw 5000 planets from the core mass, initial atmospheric mass fraction, and period distributions defined in their paper. For the core mass distribution and initial atmospheric mass fraction distribution, both defined by fifth-order Bernstein polynomials, we take the value of the coefficients to be the mean values from their MCMC posteriors. We take the core density to be the same for each planet, based on the best-fit model III posteriors. Each planet gets randomly assigned a star from the \Kepler sample. The input distributions can be seen in the top row of Figure~\ref{fig:input-dists}. All planets are evolved through 3Gyr of EUV/X-ray photoevaporation. We take their final radii and periods to be the intrinsic population for comparison and calculate each planet's insolation based on their respective stellar properties and Equation~\ref{eq:insol}.

The resulting occurrence contours can be seen in Figure~\ref{fig:FGK-comp}. Column two shows the comparison of our measured FGK occurrence and photoevaporation model planet populations. In Figure~\ref{fig:dTypes-comp} and Figure~\ref{fig:dTypes-comp-insol}, we look at different stellar type bin occurrence.
Qualitatively, the photoevaporation model has a similar spread in radii, but is flatter across orbital period. There is a marked difference when comparing F, G, and K occurrence. The model over-populates the low period sub-Neptunes for F-stars and under-populates them for K-stars. Reasons for this are discussed in Section~\ref{sec:discussion}.

We statistically compare these populations using a KS test in 1D (radius) and in 2D (period and radius). For the 1D test, we compare planet radii from 2.5-8 \Re, for periods less than 100 days, and for the 2D test, we compare the radii and periods of planets 2.5-8 \Re and with periods less than 100 days. The p-values of these tests are shown in Table~\ref{tab:RO-stats}, all of which are less than our threshold of $p < 0.05$. Therefore we can reject the null hypothesis that the planets come from the same underlying distribution. 

\begin{deluxetable}{ccc}
    \tablecaption{Photoevaporation Model Comparison}
    \tablehead{\colhead{Stellar Type} & \colhead{KS 1D} & \colhead{KS 2D} \\ 
    \colhead{} & \colhead{(p-value)} & \colhead{(p-value)} } 
    \startdata
    FGK & 0.022 & 0.0002\\
    F &  0.001  &  6.4e-15 \\
    G &  0.003  &  0.001 \\
    K &  1.9e-5  &  0.0001 \\
    \enddata

    \tablecomments{P-values for each statistical test for each stellar type group for our Model I. The second column shows p-values for a 1D KS test in radius. The last column shows p-values for a 2D KS test in radius and period. We reject the null hypothesis for all tests that the occurrence distribution is the same as the model distribution.}

    \label{tab:RO-stats}

\end{deluxetable}


\subsection{Model II: Core-Powered Mass-Loss}\label{sec:cpml}
Next, we build a core-powered mass-loss model of based on the model of \citet{gupta2020}. For the input distributions, we draw 5000 planets from the core mass and period distributions defined in their paper, each with an initial atmospheric mass fraction of $f=0.05(M_c/M_{\oplus})^{1/2}$. Each planet gets randomly assigned a star from the \Kepler sample. The input distributions are shown in row two of Figure~\ref{fig:input-dists}. All planets are evolved through 3Gyr of mass loss. We take their final radii and periods to be the intrinsic population for comparison. We calculate each planet's insolation based on their stellar properties and Equation~\ref{eq:insol}.

Figure~\ref{fig:FGK-comp} shows the comparison of our measured FGK occurrence and core-powered mass-loss model in column three. The same comparison for F, G, and K-type occurrence can also be seen in column three of Figure~\ref{fig:dTypes-comp} for period-radius space and in Figure~\ref{fig:dTypes-comp-insol} for insolation-radius space. Qualitatively, core-powered mass-loss produces a much more narrow radius distribution for the sub-Neptunes than photoevaporation does, i.e., there are fewer large ($>3$\Re) sub-Neptunes, and the edge of the sub-Neptunes have a slope that matches the radius valley.

As with our first model comparison, we statistically compare these populations using a KS test in 1D and in 2D. For the 1D test, we compare planet radii from 2.5-6 \Re, for periods less than 100 days and for the 2D test, we compare the radii and periods of planets 2.5-6 \Re and with periods less than 100 days. The p-values for these tests can be seen in Table~\ref{tab:GS-stats}. All of the statistics lie well below our threshold of $p < 0.05$, meaning we can reject the null hypothesis that the planet samples come from the same distribution.

\begin{deluxetable}{cccc}
    \tablecaption{Core-Powered Mass-Loss Model Comparison}
    \tablehead{\colhead{Stellar Type} & \colhead{KS 1D} & \colhead{KS 2D} \\ 
    \colhead{} & \colhead{(p-value)} & \colhead{(p-value)} } 
    \startdata
    FGK & 3.7e-19 & 7.7e-25\\
    F & 1.4e-25 & 2.6e-58 \\
    G & 1.2e-23 & 1.2e-30 \\
    K & 2.6e-12 & 1.1e-16 \\
    \enddata

    \tablecomments{P-values for each statistical test for each stellar type group for our Model II. The second column shows p-values for a 1D KS test in radius. The last column shows p-values for a 2D KS test in radius and period. We reject the null hypothesis for all tests that the occurrence distribution is the same as the model distribution.}

    \label{tab:GS-stats}

\end{deluxetable}

\section{Discussion}\label{sec:discussion}
Of the two models we compared against, neither were able to replicate the shape of the Neptune desert and radius cliff statistically. These models were optimized for the radius valley, so it is not surprising that they do not precisely agree with the distribution of the radius cliff.
In the case of the core-powered mass-loss model, we see that the largest sub-Neptunes have been \textit{over}-stripped, as they sit at too small a radius compared to our occurrence distribution.
Photoevaporation is a closer match, as it has the widest distribution of sub-Neptune radii. 
Because our photoevaporation model is able to keep a portion of the large sub-Neptunes 3Gyr of evolution, photoevaporation may be the dominant mass loss mechanism for sub-Neptunes larger than 2.5\Re.
This agrees with the analysis of \citet{owen2023}. They find that core-powered mass-loss dominates for less massive, closer-in planets and that photoevaporation dominates for more massive, farther-out planets. 

Further, we have tested each physical process independently of each other, but planets are likely to undergo both photoevaporation and core-powered mass-loss in their lifetimes. \citet{owen2023} use a semi-analytic model to study the transition from core-powered mass-loss to photoevaporation for close-in planets and find that most small planets will undergo both mechanisms. Even for those that do not initially undergo core-powered mass-loss, their photoevaporation can be ``enhanced'' by core-powered mass-loss.

The agreement between model and occurrence is worse when separated by stellar type.
This is somewhat expected because the photoevaporation model does not specifically vary EUV/X-ray flux, and so cannot account for changes in photodissociation that is expected to occur for the planets.

Changes in the location of the peak of the sub-Neptune radius distribution as a function of period and insolation has previously been reported. \citet{mcdonald2019} measured planet occurrence as a function of lifetime flux and found that the radius distribution for sub-Neptunes moved as a function of bolometric flux but not for X-ray flux, implying that photoevaporation (the X-ray dominant process) is the cause. The shift in radius as a function of stellar type (or stellar mass) can be seen in \citet{dattilo2023} and \citet{ho2024}, respectively.

All of our models do yield a larger population of sub-Neptunes around K-stars than earlier stellar types. This is because each planet receives less flux from the host star (EUV or bolometric) and each planet receives, and therefore loses, less of their atmospheres and remain sub-Neptunes.


\subsection{Possible Hypotheses to Explain the Measured Cliff Shape}
We have shown that neither photoevaporation nor core-powered mass loss models that reproduce the radius valley can reproduce the radius cliff within 100 days. The occurrence distributions show more large sub-Neptunes at longer orbital periods than are produced by the models. There are several possible reasons why this could occur.

A single physical process may dominate the history of the sub-Neptunes, as we have modeled here, but the initial planet distributions may be different from those that form the radius valley. Our measured radius cliff distribution requires initial conditions that produce larger sub-Neptunes at longer orbital periods. This is a plausible explanation that could be tested with the hierarchical Bayesian framework of \citet{rogers2021}.

There may be other modifications of the physics, such as a variable efficiency factor. The efficiency of atmospheric escape, $\eta$, is defined in Equation 7 in \citet{rogers2021}. This definition has $\eta$ dependent on the planet's escape velocity (i.e., core mass) but it could also depend on the planet's orbital period or insolation flux. \citet{owen2013} find that $\eta$ varies greatly over a star's lifetime and highly depends on planet mass, radius, and incident flux. In their planet simulations, the largest sub-Neptune planet radii vary as a function of separation, peaking at about 0.1~au before decreasing. Tuning the efficiency factor on insolation flux and orbital period could yield better match our occurrence distributions.

These models make the assumption that sub-Neptunes form in situ with an Earth-like rocky core and H/He primordial envelope. 
Theory predicts a large fraction of water-rich cores interior to 1~au \citep{aguichine2021, chakrabarty2023, burn2024}, which could impact the evolved sub-Neptune population.

It is possible that atmospheric mass loss processes are not the dominant evolutionary processes for the largest sub-Neptunes, and interactions between the planet's envelope and surface account for its evolution. \citet{kite2019} propose a model that is independent of the radius valley to explain both the location and the steepness of the occurrence cliff. \cite{kite2019} utilize the ``fugacity crisis'' to explain the cliff for planets at orbital periods less than 100 days. They argue that most transiting sub-Neptunes will still have hot enough cores so that there is a magma ocean interfacing with the hydrogen atmosphere. At high pressures ($>$ 1 GPa), H$_2$ becomes less compressible and makes it more soluble. The solubility of H$_2$ into magma increases as atmospheric pressure increases. Therefore, as more and more hydrogen accretes into the planet's atmosphere, the atmospheric pressure increases and more H$_2$ can be sequestered into the magma oceans. This phenomenon grows the planet's mass but limits its radius, explaining the occurrence cliff.



\section{Conclusion \& Summary}\label{sec:conclusions}
We have presented occurrence rates for \Kepler{} FGK stars for the radius cliff region of parameter space and tied it to two evolutionary theories. We compare these results to previous literature and see they have the same conclusions. We find that:

\begin{itemize}
    \item The radius cliff and Neptune desert, measured in planet occurrence-space, change their relative shape as a function of orbital period, radius, and insolation flux.
    \item When comparing the region of the radius cliff, $2.5-8$\Re and $10-100$ days, the intrinsic planet samples do not statistically agree with a photoevaporation model or a core-powered mass-loss model that are optimized for the radius valley.
    \item Because of the model mis-match, the radius cliff is not dominantly formed as a by-product of the formation of the radius valley.
\end{itemize}

While we have shown that the radius cliff is not produced by these specific models and initial conditions, it could be produced by the same physics with different initial conditions. The differences between models, and domain locations where theoretical models are a poor match for the observed distributions can be further tested in the mass-radius plane, where planet compositions are revealed more easily. This work is reserved for a future paper.



\begin{acknowledgments}
    \begin{center}
        ACKNOWLEDGEMENTS
    \end{center}
    We thank the referee for providing helpful comments that improved the manuscript.
    AD would like to thank James Rogers, James Owen, Ruth Murray-Clay, Michelle Kunimoto, and Eric Lopez for helpful conversations in the production of this manuscript. 
    AD gratefully acknowledges support from the Heising-Simons Foundation through grant 2021-3197.
    This material is based upon work supported by NASA’S Interdisciplinary Consortia for Astrobiology Research (NNH19ZDA001N-ICAR) under award number 19-ICAR19 2-0041.
    We acknowledge use of the lux supercomputer at UC Santa Cruz, funded by NSF MRI grant AST 1828315.
    This work used the Extreme Science and Engineering Discovery Environment (XSEDE) \textit{Expanse} at the San Diego Supercomputer Center through allocation TG-PHY210033 \citep{xsede}.
\end{acknowledgments}

\software{\textit{Astropy} \citep{astropy}, \textit{numpy} \citep{numpy}, \textit{pandas} \citep{pandas}, \textit{scipy} \citep{scipy}, \textit{matplotlib} \citep{matplotlib}}


\bibliography{sample631}{}
\bibliographystyle{aasjournal}



\end{document}